\documentclass[twocolumn,floatfix,amsmath,nofootinbib,amssymb,preprintnumbers,aps,showkeys]{revtex4}
\usepackage{graphicx}
\usepackage{booktabs}
\usepackage{capt-of}
\usepackage{bm}
\usepackage{txfonts}
\usepackage{slashed}
\usepackage{xcolor}
\usepackage{array}
\usepackage[normalem]{ulem}
\usepackage{CJKutf8}
\usepackage[colorlinks=true,citecolor=blue,linkcolor=red,urlcolor=blue]{hyperref}
\begin{document}
\begin{CJK*}{UTF8}{gbsn}
\title{Phase-constrained $\Sigma^*$ spectroscopy in the pure-$I=1$ reactions
$K_Lp\to\pi^+\Sigma^0$ and $K_Lp\to\pi^+\Lambda$}

\author{Dan Guo(郭丹)$^{1}$}\email{guod13@ysu.edu.cn}
\author{Marshall B. C. Scott$^{2}$}\email{marshall.scott@gwu.edu}
\author{Igor Strakovsky$^{2}$}\email{igor@gwu.edu}
\author{Fu-Rong Xu(许甫荣)$^{3,5}$}\email{frxu@pku.edu.cn}
\author{Bing-Song Zou(邹冰松)$^{4}$}\email{zoubs@mail.tsinghua.edu.cn}

\affiliation{
$^1$Key Laboratory for Microstructural Material Physics of Hebei Province,
School of Science, Yanshan University, Qinhuangdao 066004, China\\
$^2$Institute for Nuclear Studies, Department of Physics,
The George Washington University, Washington, DC 20052, USA\\
$^3$School of Physics and State Key Laboratory of Nuclear Physics and Technology,
Peking University, Beijing 100871, China\\
$^4$Department of Physics and Center for High Enegy Physics, Tsinghua University, Beijing 100084, China\\
$^5$Southern Center for Nuclear-Science Theory (SCNT),
Institute of Modern Physics, Chinese Academy of Sciences, Huizhou 516000, China
}

\begin{abstract}
Low-energy $\bar K N$ scattering provides direct access to strange-baryon spectroscopy, but conventional charged-kaon reactions mix the $I=0$ and $I=1$ amplitudes. 
Motivated by the isospin-selective nature of the $K_Lp$ reactions, we present, to our knowledge, the first simultaneous analysis of the pure-$I=1$ reactions $K_Lp\to\pi^+\Sigma^0$ and $K_Lp\to\pi^+\Lambda$ in which a common set of resonance parameters and a fixed relative-phase convention are imposed on both final states. 
Within an effective-Lagrangian model, a joint fit of available differential cross sections and recoil polarizations yields $\chi^2/\mathrm{d.o.f.}=1.604$. 
We find clear channel complementarity: the $t$-channel $K^*$ exchange is more important in $K_L p\to \pi^+\Lambda$, especially at forward angles, whereas $K_L p\to \pi^+\Sigma^0$ is more sensitive to the contribution of $\Sigma(1620)\,1/2^-$. At the same time, $\Sigma(1660)\,1/2^+$ remains important through interference effects in both channels. These results demonstrate the importance of multichannel, phase-constrained analyses for establishing the $I=1$ hyperon spectrum and provide timely phenomenological input for future high-precision measurements by the KLF program at JLab. Once the pure $I=1$ amplitudes are reliably determined, they will also enable the separation of the $I=0$ component in the much more abundant $K^- p\to \pi^\pm\Sigma^\mp$ data, opening a path toward a more quantitative study of the $\Lambda^*$ spectrum. 
\end{abstract}

\keywords{$\bar{K}N$ scattering; hyperon spectroscopy;
isospin-selective reactions; $\Sigma$ resonances; phase convention}

\maketitle
\end{CJK*}
\section{Introduction}

Hadron scattering provides one of the most direct windows on the nonperturbative dynamics of QCD. Among hadronic reactions, $\pi N$ scattering has historically been the primary benchmark for baryon spectroscopy~\cite{Klempt:2009pi,Crede:2013kia}. Most established $N^*$ and $\Delta^*$ resonances have been constrained crucially by $\pi N$ data~\cite{ParticleDataGroup:2024cfk}, and basic hadronic quantities such as the $\pi N\bar N$, $\rho N\bar N$ couplings are also determined most directly in this channel~\cite{Arndt:2006bf, Briscoe:2023gmb,Julia-Diaz:2007qtz,Matsuyama:2006rp, Ireland:2019uwn}. 
In this sense, $\pi N$ scattering has long served as the benchmark process for studying the excitation spectrum of nonstrange baryons.

A closely analogous role in strange baryon spectroscopy is played by low-energy $\bar{K}N$ scattering~\cite{Kamano:2014zba, Kamano:2015hxa, Oller:2000fj, Oller:2006jw, Oset:1997it, Khemchandani:2018amu, Zhang:2013sva}. In contrast to many other hadronic reactions, the $\bar{K}N\to \pi\Sigma$ and $\bar{K}N\to \pi\Lambda$ channels are exothermic, since the total mass of the initial $\bar{K}N$ system is larger than those of the corresponding final states. Although a physical scattering process is realized at finite beam momentum, this kinematic feature implies that these channels remain open already in the zero-momentum limit and therefore couple very strongly near threshold. As a consequence, low-energy $\bar{K}N$ scattering is characterized by mb-scale hadronic cross sections and strong couplings to the $\pi\Sigma$ and $\pi\Lambda$ channels, making it one of the most favorable arenas for investigating hyperon resonances and strange baryon dynamics. 

This opportunity has motivated extensive studies of the $\bar{K}N$ system over many years~\cite{Kamano:2014zba, Kamano:2015hxa, Oller:2000fj, Oller:2006jw, Oset:1997it, Khemchandani:2018amu, Zhang:2013sva}. Most of the available hyperon-scattering data have been accumulated in $K^- p$ reactions and analyzed in partial-wave and coupled-channel approaches~\cite{SAID,BnGaHyperon, Feijoo:2018den, Feijoo:2021zau}. The strong coupled-channel nature of the $S=-1$ sector and the presence of nearby thresholds make the extraction of resonance parameters particularly challenging~\cite{Lu:2022hwm}. Recent progress has shown that cross-channel information can substantially reduce the uncertainties of low-energy meson-baryon amplitudes, while the low-lying $\Sigma^*$ and $\Lambda^*$ spectra, especially the long-debated $\Sigma(1/2^-)$ sector, still remain far less settled than the nonstrange baryon spectrum~\cite{Lu:2022hwm,Wang:2024jyk,Ma:2025wbz,Lyu:2024qgc}. One important reason is that the conventional $K^- p$ reactions generally contain both $I=0$ and $I=1$ contributions, so that the resonance content of a given isospin channel is not easily isolated. 

This is precisely where $K_L p$ scattering becomes uniquely valuable. In particular, the reactions $K_L p\to \pi^+\Sigma^0$ and $K_L p\to \pi^+\Lambda$ isolate the pure $I=1$ amplitudes of $\bar{K}N\to \pi\Sigma$ and $\bar{K}N\to \pi\Lambda$, respectively. Therefore, they provide exceptionally clean probes of the $\Sigma^*$ spectrum and, more generally, of the isovector $\bar{K}N$ interaction \cite{Amaryan:2016ufk,GlueX:2017hgs,Guo:2025mha}. This opportunity is particularly timely for the K-Long Facility (KLF) at Jefferson Lab, which is designed to provide high-statistics measurements with broad angular and energy coverage~\cite{GlueX:2017hgs,KLF:2020gai}. 

The spectroscopy of low-lying strange baryons remains an active subject from both phenomenological and theoretical perspectives. In particular, the existence, mass, and width of the low-lying $\Sigma(1/2^-)$ states are still under debate \cite{Wang:2024jyk,Ma:2025wbz}. Our previous study of the isospin-selective reaction $K_L p\to \pi^+\Sigma^0$ has shown that such channels provide valuable constraints on the $\Sigma^*$ spectrum when differential cross sections and recoil polarizations are analyzed within a unified phase convention \cite{Guo:2025mha}. In parallel, earlier analyses of the $\pi\Lambda$ channel have highlighted the importance of interference effects and polarization observables for extracting the properties of $\Sigma$ resonances \cite{Gao:2010ve,Gao:2012zh}. These developments strongly suggest that a combined analysis of multiple isospin-selective $K_L p$ channels can provide more stringent constraints than any single channel alone. 
Following the PDG convention, we specify relative phases by the signs of the imaginary parts of the resonant partial-wave amplitudes at resonance. 

\section{Framework and fit strategy}\label{sec:framework}

\begin{figure}[!htbp]
    \centering
    \includegraphics[width=\columnwidth]{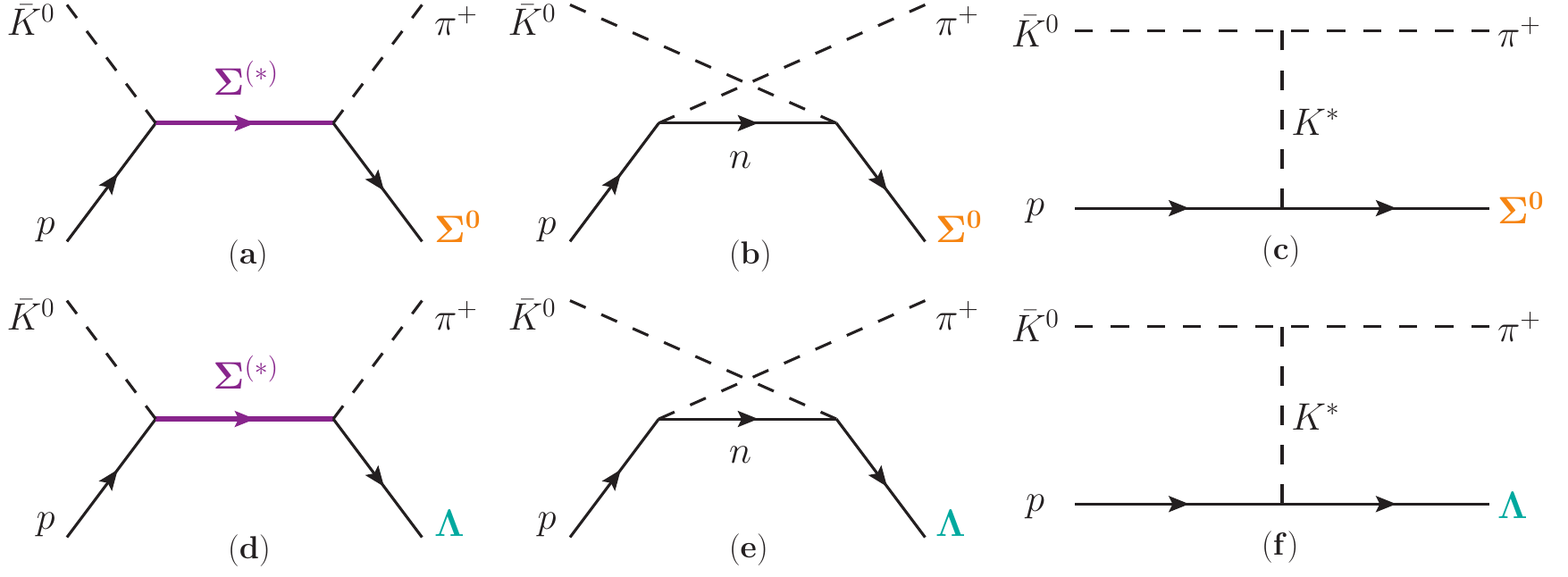}
    \caption{Tree-level mechanisms for $\bar K^0(k_1)p(p_1)\to\pi^+(k_2)\Sigma^0(p_2)$ and $\bar K^0(k_1)p(p_1)\to\pi^+(k_2)\Lambda(p_2)$. Panels (a,d), (b,e), and (c,f) denote, respectively, $s$-channel $\Sigma^{(\ast)}$ exchange, $u$-channel nucleon exchange, and $t$-channel $K^\ast$ exchange.}
    \label{fig:feynman}
\end{figure}

The nearly fourfold longer lifetime of the $K_L$ meson relative to the $K^-$ ($51.16$ versus $12.38$ ns) extends $K_Lp$ measurements to lower incident momenta, providing a valuable complement to $K^-p$ scattering~\cite{ParticleDataGroup:2024cfk}. The tagged $K_L$ beam at KLF will further enable precision measurements with event-by-event reconstruction of the incident momentum. Neglecting $CP$ violation at the present level of accuracy, we treat $K_L$ and $K_S$ as $CP$ eigenstates:
$K_L=\frac{1}{\sqrt{2}}(K^0-\bar{K}^0), K_S=\frac{1}{\sqrt{2}}(K^0+\bar{K}^0).$

Strangeness conservation selects the $\bar K^0p$ component in the reactions of interest.  Their amplitudes are therefore pure isovector combinations,
\begin{align}
    T(K_Lp\to\pi^+\Sigma^0)&=-\frac{1}{2} T^1(\bar KN\to\pi\Sigma),\nonumber\\
    T(K_Lp\to\pi^+\Lambda)&=-\frac{1}{\sqrt{2}} T^1(\bar KN\to\pi\Lambda).
    \label{eq:pureI1}
\end{align}
In contrast, the charged $K^-p\to\pi^\pm\Sigma^\mp$ reactions contain both
$I=0$ and $I=1$ amplitudes~\cite{Amaryan:2016ufk}.
\begin{align}
    T(K^-p\to\pi^+\Sigma^-)&=\frac{1}{2}T^1(\bar{K}N\to\pi\Sigma)-\frac{1}{\sqrt{6}}T^0(\bar{K}N\to\pi\Sigma),\nonumber\\
    T(K^-p\to\pi^-\Sigma^+)&=-\frac{1}{2}T^1(\bar{K}N\to\pi\Sigma)-\frac{1}{\sqrt{6}}T^0(\bar{K}N\to\pi\Sigma).
\end{align}

The model contains the three mechanisms shown in Fig.~\ref{fig:feynman}.  The $s$ channel includes the established four-star $\Sigma(1189)\,1/2^+$, $\Sigma(1385)\,3/2^+$, $\Sigma(1670)\,3/2^-$, and $\Sigma(1775)\,5/2^-$ states, together with the candidate $\Sigma(1580)\,3/2^-$, $\Sigma(1620)\,1/2^-$, $\Sigma(1660)\,1/2^+$, and $\Sigma(1750)\,1/2^-$ states.  A dipole form factor is attached to each hadronic vertex, and the couplings and cutoff parameters are restricted to ranges motivated by SU(3) relations~\cite{Gao:2010ve, Kamano:2013iva, Oh:2004wp, Doring:2010ap, Oh:2007jd, Oh:2006hm, Arndt:2006bf, Mueller-Groeling:1990uxr} and decay-width ranges compiled by the Particle Data Group (PDG)~\cite{ParticleDataGroup:2024cfk}. The complete effective Lagrangians~\cite{Gao:2010ve, Gao:2012zh, Kamano:2013iva}, coupling conventions~\cite{Stoks:1999bz, Oh:2006hm, Oh:2004wp, Shi:2014vha}, propagators, amplitudes, and definitions of the differential cross section and recoil polarization~\cite{Amaryan:2016ufk, Shi:2014vha, Penner:2002ma} are given in \hyperref[supplementary]{Supplementary materials} S1 and S2.

We fit 529 differential-cross-section and 38 recoil-polarization measurements from the historical $K_Lp$ data sets~\cite{
Bologna-Edinburgh-Glasgow-Pisa-Rutherford:1977cyz,Cho:1975dv,
Engler:1978rr,Burkhardt:1975sd,Corden:1979tz}.  The same 34 adjustable parameters are used for both final states.  The KLF Monte Carlo points shown below are prospective projections and are not included in the minimization.

\section{Results and discussion}\label{sec:results}

\renewcommand{\arraystretch}{1.5}  
\begin{table}[tbph]
    \centering
    \caption{Optimal fit parameters for less-established $\Sigma^*$ resonances. The complete parameter set and alternative fits are listed in Table S1 of Supplementary materials. The partial widths are estimated from the vertex couplings, and the $\Gamma_{ tot}$ is obtained from the fit. }
    \label{tab:optimal}
    \setlength{\tabcolsep}{8pt}
    \begin{tabular}{l|ll}
    \toprule[1.50pt]
    \toprule[0.50pt]
    Resonances & Parameters & Optimal Fit \\ \hline
    $\Sigma(1580)\,3/2^-$ & $\sqrt{\Gamma_{\bar{K}N}\Gamma_{\pi\Sigma} }/\Gamma_{tot}$ & $+0.010\pm0.004$ \\
                         & $\sqrt{\Gamma_{\bar{K}N}\Gamma_{\pi\Lambda} }/\Gamma_{tot}$ & $-0.012\pm0.002$ \\
                         \hline
    $\Sigma(1660)\,1/2^+$ & $M$ [GeV] & $1.637\pm0.003$ \\
                         & $\Gamma$ [GeV] & $0.130\pm0.006$ \\
                         & $\sqrt{\Gamma_{\bar{K}N}\Gamma_{\pi\Sigma} }/\Gamma_{tot}$ & $-0.052\pm0.005$ \\
                         & $\sqrt{\Gamma_{\bar{K}N}\Gamma_{\pi\Lambda} }/\Gamma_{tot}$ & $-0.077\pm0.002$ \\
                         \hline
    $\Sigma(1620)\,1/2^-$ & $M$ [GeV] & $1.557\pm0.002$ \\
                         & $\Gamma$ [GeV] & $0.117\pm0.001$ \\
                         & $\sqrt{\Gamma_{\bar{K}N}\Gamma_{\pi\Sigma} }/\Gamma_{tot}$ & $-0.741\pm0.006$ \\
                         & $\sqrt{\Gamma_{\bar{K}N}\Gamma_{\pi\Lambda} }/\Gamma_{tot}$ & $-0.142\pm0.006$ \\
                         \hline
    $\Sigma(1750)\,1/2^-$ & $\sqrt{\Gamma_{\bar{K}N}\Gamma_{\pi\Sigma} }/\Gamma_{tot}$ & $+0.45\pm0.01$ \\
                         & $\sqrt{\Gamma_{\bar{K}N}\Gamma_{\pi\Lambda} }/\Gamma_{tot}$ & $+0.19\pm0.01$ \\
                         \hline
            & \text{D.o.F}           & 533  \\
            & $\chi^2/\text{D.o.F}$  & 1.604 \\
    \bottomrule[0.50pt]
    \bottomrule[1.50pt]
    \end{tabular}
\end{table}
\setlength{\extrarowheight}{0pt}

\begin{figure*}[!htbp]
    \centering
    \includegraphics[width=0.85\textwidth]{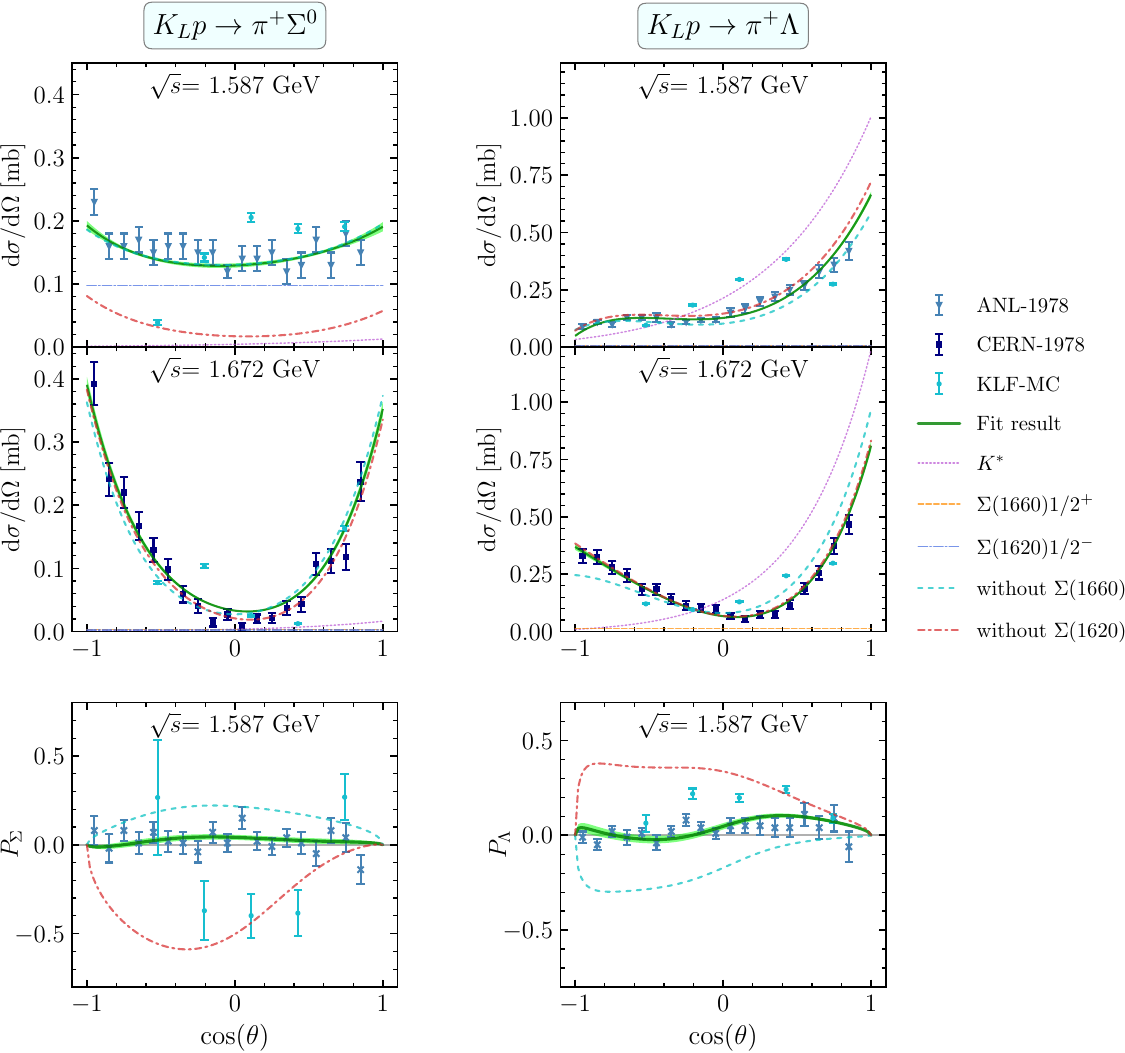}
    \caption{Selected differential cross sections (upper two rows) and recoil polarizations (bottom row) for $K_Lp\to\pi^+\Sigma^0$ (left) and $K_Lp\to\pi^+\Lambda$ (right), with 1$\sigma$ uncertainty bands.  Cross sections are shown at $\sqrt{s}=1.587$ and 1.672~GeV, and polarizations at 1.587~GeV. The full fit, isolated $K^\ast$, $\Sigma(1660)$, and $\Sigma(1620)$ contributions, and results obtained after removing either resonance are compared with ANL-1978~\cite{Engler:1978rr} and CERN-1978~\cite{Bologna-Edinburgh-Glasgow-Pisa-Rutherford:1977cyz} data. KLF Monte Carlo points are prospective projections and were not included in the fit.}
    \label{fig:selected}
\end{figure*}

\begin{figure*}[!htbp]
    \centering
    \includegraphics[width=0.95\textwidth]{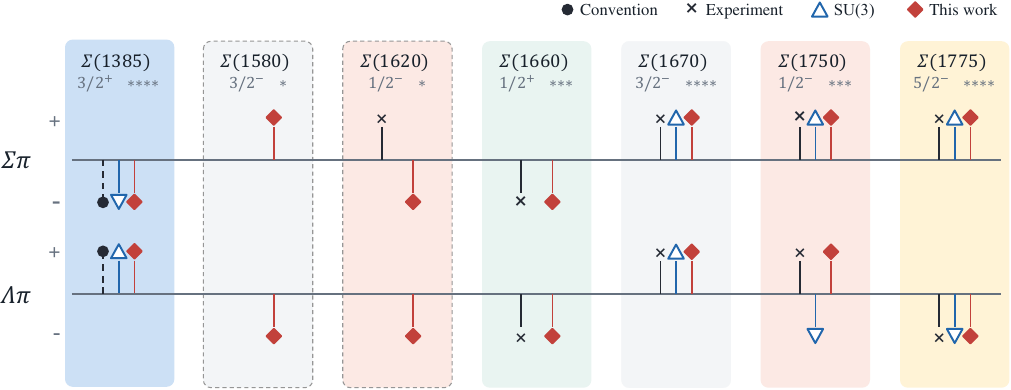}
    \caption{Relative signs of the $\Sigma^\ast$ couplings to $\bar KN\to\Sigma\pi$ (upper row) and $\bar KN\to\Lambda\pi$ (lower row). Black circles define the common convention, crosses denote experimental assignments, open triangles show SU(3) expectations, and red diamonds are the optimal-fit signs.  
    The dashed outlines mark one-star candidates.}
    \label{fig:phases}
\end{figure*}

Earlier single-channel analyses established the sensitivity of the $\pi\Sigma$ reaction~\cite{Guo:2025mha} to a low-lying $\Sigma(1/2^-)$ contribution and the importance of interference in the $\pi\Lambda$ channel~\cite{Gao:2010ve}. They did not, however, exploit two final states with distinct background mechanisms and polarization patterns under common parameter and phase constraints. 


The optimal fit describes 567 data points with 34 parameters, giving $\chi^2/\mathrm{d.o.f.}=1.604$ for 533 degrees of freedom, shown in Table~\ref{tab:optimal}.  A baseline fit containing only the nonresonant terms and the established four-star states
gives $\chi^2/\mathrm{d.o.f.}=5.161$. These contributions are always retained in the fit, and attempts to remove any of them lead to a visibly worse description of the data. 
Alternative tests summarized in Table S1 of Supplementary materials show that removing $\Sigma(1750)$ raises this value to 2.210, indicating that this state becomes important once the two channels are analyzed simultaneously, even though its role was not strongly required in our previous single-channel study~\cite{Guo:2025mha}.
When fixing the $\Sigma(1620)$ mass at 1.4~GeV gives $\chi^2/\mathrm{D.o.F.}$=1.880, and drives the fitted $\Sigma(1660)\,1/2^+$ mass and width toward the edges of the allowed ranges, showing that such a low-mass assignment for the $\Sigma(1620)$ is disfavored by the present combined data.
And adding $\Sigma(1620)$ alone to the baseline already reduces it to 2.546.  Thus the low-lying $1/2^-$ contribution accounts for an essential part of the missing strength, while the global description also requires the interference supplied by the other candidate states. 

In optimal fit, for the $\Sigma(1660)\,1/2^+$, we obtain Breit-Wigner $M=1.637\pm0.003$ GeV and $\Gamma=0.130\pm0.006$ GeV, in good agreement with our previous analyses of the $\pi\Lambda$ channel~\cite{Gao:2010ve,Gao:2012zh}. For the $\Sigma(1620)\,1/2^-$, the optimal fit gives $M=1.557\pm0.002$ GeV and $\Gamma=0.117\pm0.001$ GeV. 
Within the present parametrization the fitted data favor an $S$-wave contribution near 1.55~GeV. They do not resolve the nature of the structure at the $\bar KN$ threshold~\cite{Belle:2022ywa}.

Bonn--Gatchina questioned whether its narrow $1/2^-$ state near 1.68~GeV could be identified with $\Sigma(1620)$~\cite{Matveev:2019igl,Sarantsev:2019xxm}, while JPAC found no evidence for $\Sigma(1620)$ in its fit to the Kent State single-energy amplitudes~\cite{Fernandez-Ramirez:2015tfa}. In a unitarized chiral fit including $s$ and $p$ waves, Feijoo \textit{et al.} obtained a broad isovector $1/2^-$ pole near 1.59~GeV whose width and couplings disfavor that assignment~\cite{Feijoo:2021zau}. 
An automated resonance-selection scheme for hadron spectroscopy was proposed in Ref.~\cite{Landay:2018wgf}. 
Table~\ref{tab:sigma-pwa} compares the two isovector partial waves. In $S_{11}$, the real part of the ANL--Osaka Model~B pole (1551~MeV) lies close to our Breit--Wigner mass (1557~MeV), although its full pole width of 376~MeV exceeds our fitted width of 117~MeV. In $P_{11}$, the present fit and BnGa give similar Breit--Wigner masses (1637 and 1665~MeV) and different widths (130 and 300~MeV).

\begin{table}[tbph]
    \centering
    \caption{Selected isovector $S_{11}$ and $P_{11}$ resonance parameters (central values in MeV). Pairs give Breit--Wigner $(M_{\rm BW},\Gamma_{\rm BW})$ for this work and Bonn--Gatchina (BnGa)~\cite{Sarantsev:2019xxm}. ANL--Osaka poles are shown for $1400<\mathrm{Re}\,W_p/\mathrm{MeV}<1750$ ~\cite{Kamano:2015hxa}.}
    \label{tab:sigma-pwa}
    \renewcommand{\arraystretch}{1.5}
    \setlength{\tabcolsep}{8pt}
    \begin{tabular}{l|ll}
        \toprule[1.50pt]
        \toprule[0.50pt]
        Analysis & $S_{11}$ $(1/2^-)$ & $P_{11}$ $(1/2^+)$ \\ \hline
        This work & $(1557,117)$ & $(1637,130)$ \\
        BnGa      & $(1681,40)$  & $(1665,300)$ \\ \hline
        ANL--Osaka A & $1704-43i$ &
            {\renewcommand{\arraystretch}{1.15}$\begin{array}{@{}l@{}}1547-92i\\1706-51i\end{array}$} \\ \hline
        ANL--Osaka B & $1551-188i$ &
            {\renewcommand{\arraystretch}{1.15}$\begin{array}{@{}l@{}}1457-39i\\1605-96i\end{array}$} \\
        \bottomrule[0.50pt]
        \bottomrule[1.50pt]
    \end{tabular}
\end{table}

The statistical bands produced by the combined fit are narrow in all displayed observables.  Within the adopted tree-level model, this indicates that imposing one parameter set on two complementary final states substantially reduces the fit correlations found in single-channel studies. 
The bands represent propagated fit uncertainties conditional on the adopted tree-level amplitude, resonance content, parameter ranges, and statistical treatment of the data. They do not include model-systematic uncertainties associated with alternative background parameterizations, coupled-channel dynamics, or resonance content. Their narrowness therefore indicates strong constraints within the adopted model rather than model-independent precision. 
The very small parameter uncertainties are not only substantially better than that of the baseline fit including only the nonresonant background and the established four-star states, but are also fully consistent with the nearly invisible uncertainty bands of differential cross sections and polarizations completely shown in \hyperref[supplementary]{Supplementary materials}. These features indicate that the simultaneous fit to the $\pi^+\Sigma^0$ and $\pi^+\Lambda$ channels imposes strong constraints on the common amplitudes and significantly reduces the residual freedom in the parameter space. 
The differential cross sections and polarizations for $K_Lp\to\pi^+\Sigma^0$ and $K_Lp\to\pi^+\Lambda$ at selected $\sqrt{s}$ are plotted in Fig.~\ref{fig:selected}. 

Figure~\ref{fig:selected} exposes the complementarity that is obscured when the two reactions are studied separately.  
In $\pi^+\Lambda$, $K^\ast$ exchange largely controls the magnitude and forward-angle rise; its isolated contribution exceeds the full result at forward angles, revealing destructive interference with the remaining amplitudes.  The direct $\Sigma(1620)\,1/2^-$ contribution is much larger in $\pi^+\Sigma^0$, and removing it changes both the low-energy angular distribution and the recoil polarization.  Its effect on the $\pi^+\Lambda$ cross section is less direct, but that channel restricts the same coupling and phase through a different interference pattern.  The contribution of $\Sigma(1660)\,1/2^+$ is modest in isolation yet its removal alters both channels, especially their polarizations.  
The combined fit therefore converts the apparent single-channel difference into a quantitative statement of reaction-channel dependence.

\begin{figure}[!htbp]
  \centering
  \begin{tabular}{l}
    \includegraphics[width=\columnwidth]{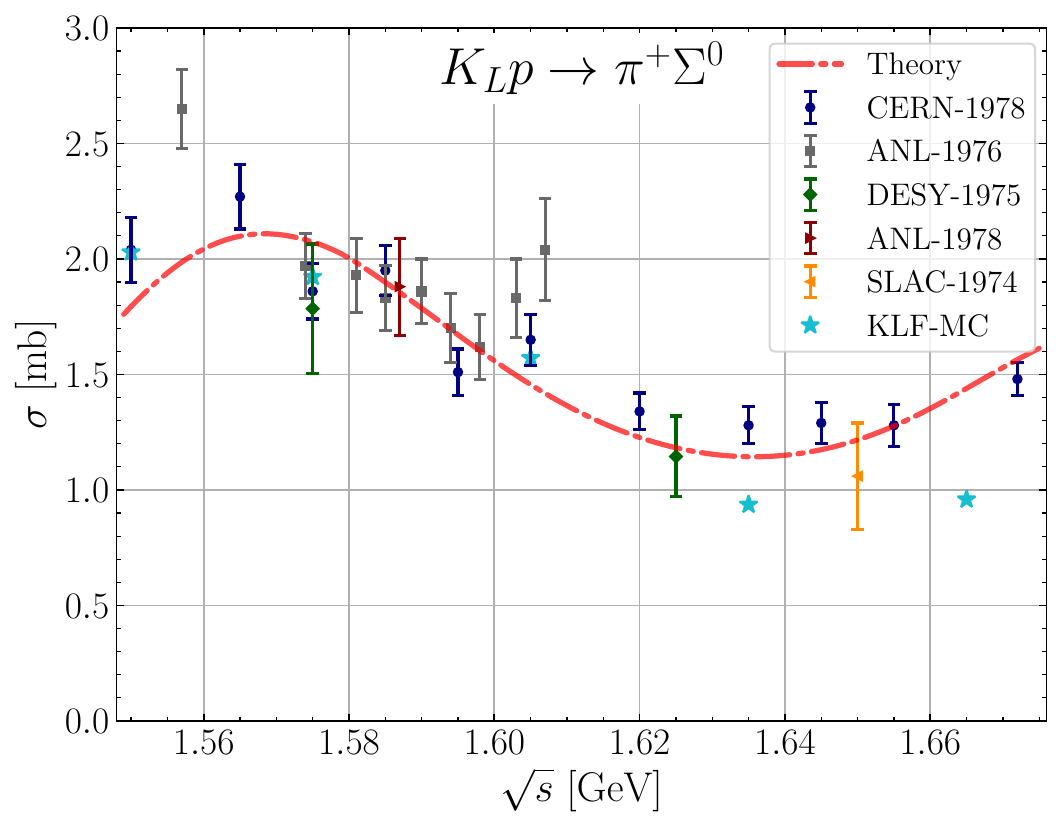} \\
    \includegraphics[width=\columnwidth]{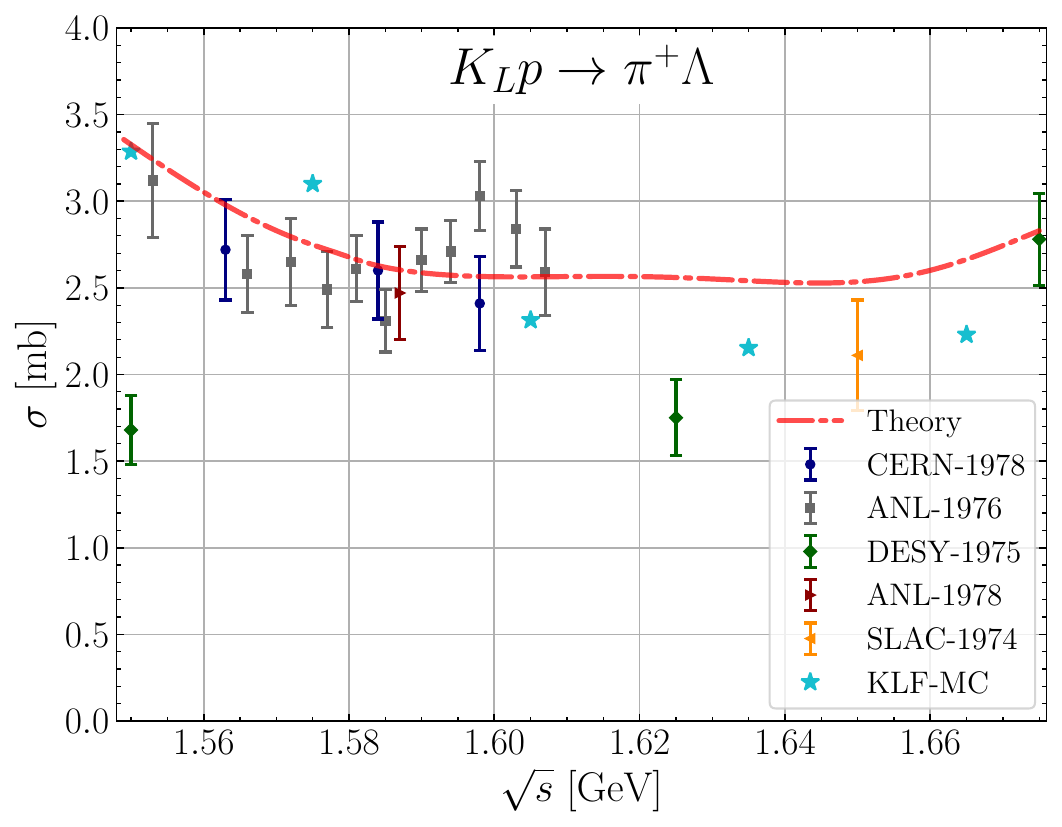} \\
  \end{tabular}
  \caption{The comparison of predicted $K_Lp\to\pi^+\Sigma^0$ and $\pi^+\Lambda$ total cross section with experimental data from CERN-1978~\cite{Bologna-Edinburgh-Glasgow-Pisa-Rutherford:1977cyz}, ANL-1976~\cite{Engler:1976cn}, DESY-1975~\cite{Burkhardt:1975sd}, ANL-1978~\cite{Engler:1978rr}, SLAC-1974~\cite{Yamartino:1974sm} and KLF Monte Carlo simulations.}\label{fig:tot}
\end{figure}

The simultaneous description of the $\pi^+\Sigma^0$ and $\pi^+\Lambda$ channels requires one common set of resonance parameters and relative phases to account for both reactions at the same time. This provides a significantly stronger restriction on the allowed interference pattern and therefore on the extracted resonance content.
The phase convention adopted here is the same as that used in our previous studies~\cite{Guo:2025mha,Gao:2010ve,Gao:2012zh,Shi:2014vha} and is consistent with the convention summarized in the PDG review, ``\textit{$\Lambda$ and $\Sigma$ Resonances}"~\cite{ParticleDataGroup:2024cfk}.

Although an individual coupling sign is convention dependent, relative signs affect observable interference once the external-state and vertex phases are fixed consistently. Figure~\ref{fig:phases} compares all fitted, experimental, and SU(3) signs only after they have been mapped to the same convention. 
To our knowledge, this is the first analysis in which one fixed phase convention is imposed simultaneously on these two independent pure-$I=1$ channels.  
The fitted signs agree with the available assignments for the established $\Sigma(1385)$, $\Sigma(1660)$, $\Sigma(1670)$, and $\Sigma(1775)$ states.  Two qualifications are important.  For $\Sigma(1620)\to\Sigma\pi$, the fitted sign is opposite to the quoted experimental assignment; because this is a one-star state with poorly determined decay information, the discrepancy is not yet decisive.  For $\Sigma(1750)\to\Lambda\pi$, the fitted sign is opposite to the SU(3) expectation but agrees with the experimental assignment~\cite{ParticleDataGroup:2024cfk}.  
The overall pattern shows that a single fixed convention permits a mutually consistent description of both reactions and most external sign assignments. 

The same parameter set also follows the overall energy dependence of both $K_Lp\to\pi^+\Sigma^0$ and $\pi^+\Lambda$ integrated cross sections, shown in Fig.~\ref{fig:tot}. 
Although the total-cross-section measurements were not included in the minimization, the optimal solution reproduces their overall energy dependence in both channels. 
The deviations among some historical measurements and the projected KLF precision further emphasize the need for modern, internally consistent data.

\section{Conclusions}\label{sec:conclusions}


Motivated by the isospin-selective nature of the $K_L p$ reaction, we have performed the first simultaneous phase-constrained analysis of $K_Lp\to\pi^+\Sigma^0$ and $K_Lp\to\pi^+\Lambda$, two complementary reactions that isolate the pure-$I=1$ $\bar{K}N$ amplitudes. 
We jointly analyze 529 differential-cross-section and 38 recoil-polarization measurements, obtaining $\chi^2/\mathrm{d.o.f.}=1.604$ with a common resonance parameter set and fixed phase convention. 
A single parameter set describes their differential cross sections, recoil polarizations, and total cross sections.  Within the adopted effective Lagrangian approach, the combined data substantially reduce statistical fit correlations relative to single-channel analyses. 

The clearest new spectroscopy result is the reaction-channel dependence of $\Sigma(1620)\,1/2^-$.  The $\pi^+\Sigma^0$ data drive the need for this state, whereas the $\pi^+\Lambda$ data provide complementary restrictions on its coupling and relative phase.  The $\Sigma(1660)\,1/2^+$ contribution remains important through interference in both channels, and the stronger $K^\ast$ exchange in $\pi^+\Lambda$ explains much of their different angular behavior.  At the same time, the common phase convention succeeds for most available resonance assignments and isolates two well-defined exceptions that future measurements can test.

High-precision angular distributions, broader coverage in energy extending down to 1.49~GeV and, especially, recoil polarizations from KLF will be decisive for separating model dependence from the present fit uncertainties.  
Once the pure $I=1$ amplitudes are fixed more accurately, they can also be combined with the abundant charged-kaon data to isolate the $I=0$ component and sharpen studies of the $\Lambda^*$ spectrum. 

\section*{Acknowledgments}

D.~G. is supported by the Yanshan University Research Start-up Fund No.~8190891. 
M.~S. and I.~S. are supported in part by the U.S. Department of Energy, Office of Science, Office of Nuclear Physics, under Award No.~DE--SC0016583. FRX is supported by the National Natural Science Foundation of China under Grants Nos. 12335007 and 12535008.

\clearpage
\setcounter{equation}{0}
\setcounter{figure}{0}
\setcounter{table}{0}
\renewcommand{\theequation}{S\arabic{equation}}
\renewcommand{\thefigure}{S\arabic{figure}}
\renewcommand{\thetable}{S\arabic{table}}
\renewcommand{\theHequation}{S\arabic{equation}}
\renewcommand{\theHfigure}{S\arabic{figure}}
\renewcommand{\theHtable}{S\arabic{table}}

\makeatletter
\title@column{%
\begin{center} \phantomsection\label{supplementary}
{\Large\bfseries Supplementary Material}\\[0.6em]
{\large\itshape Phase-constrained $\Sigma^*$ spectroscopy in the pure-$I=1$
reactions $K_Lp\to\pi^+\Sigma^0$ and $K_Lp\to\pi^+\Lambda$}\\[0.8em]
Dan Guo, Marshall Scott, Igor Strakovsky, Fu-Rong Xu, and Bing-Song Zou
\end{center}
}
\makeatother
\section*{S1: Effective Lagrangians and coupling constraints}\label{sec:S1}

This note gives the interaction vertices and parameter restrictions used in
the simultaneous analysis.  We first collect the vertices common to the
$\pi^+\Sigma^0$ and $\pi^+\Lambda$ final states and then list the
$u$- and $t$-channel interactions.

\subsection{Common s-channel vertices}

For the ground-state $\Sigma(1189)$ and the $J^P=1/2^+$ sector, we use
\cite{Gao:2010ve,Gao:2012zh,Kamano:2013iva}
\begin{align}
    \mathcal{L}_{KN\Sigma}&=\frac{g_{KN\Sigma}}{M_N+M_\Sigma}\partial_\mu\bar K\,\bar\Sigma\cdot\tau\gamma^\mu\gamma_5N +\mathrm{H.c.},\label{S-eq:KNSigma}\\
    \mathcal{L}_{\pi\Sigma\Sigma}&=i\frac{f_{\pi\Sigma\Sigma}}{m_\pi}\bar\Sigma\gamma^\mu\gamma_5\times
    \Sigma\cdot\partial_\mu\pi+\mathrm{H.c.},\\
\mathcal{L}_{\pi\Lambda\Sigma}
 &=\frac{g_{\pi\Lambda\Sigma}}{M_\Lambda+M_\Sigma}
 \bar\Lambda\gamma^\mu\gamma_5\partial_\mu\pi\cdot\Sigma
 +\mathrm{H.c.}.
\end{align}
The isospin fields are
\begin{align}
    &\bar{K} =(K^-,\bar{K}^0), \quad N=\begin{pmatrix}p\\n\end{pmatrix},\\
    &\bar{\Sigma}\cdot\tau = \begin{pmatrix}\bar{\Sigma}^0 & \sqrt{2}\,\bar{\Sigma}^+ \\\sqrt{2}\,\bar{\Sigma}^- & -\bar{\Sigma}^0 \end{pmatrix},\\
    &\pi=\left(\frac{1}{\sqrt{2}}(\pi^+ +\pi^-),  \frac{i}{\sqrt{2}}(\pi^+ -\pi^-), \pi^0\right),\\
    &\Sigma=\left(\frac{1}{\sqrt{2}}(\Sigma^+ +\Sigma^-),  \frac{i}{\sqrt{2}}(\Sigma^+ -\Sigma^-), \Sigma^0\right).
\end{align}
In Refs.~\cite{Gao:2010ve, Kamano:2013iva, Oh:2004wp}, the central coupling constants are estimated from SU(3) flavor symmetry, 
\begin{align}
    g_{KN\Sigma}=3.58,\qquad g_{\pi\Lambda\Sigma}=9.72,
\end{align}
and
\begin{equation}
    f_{\pi\Sigma\Sigma}=2(-1+\alpha)f_{\pi NN},\;
    f_{\pi NN}=\sqrt{4\pi\times0.08},\; \alpha=0.635.
\end{equation}
To account for possible SU(3)-symmetry breaking effects, we multiply the central values of
$g_{KN\Sigma}\cdot f_{\pi\Sigma\Sigma}$ and $g_{KN\Sigma}\cdot g_{\pi\Lambda\Sigma}$
by an overall factor varying from $1/2$ to 2. This allows for a broader exploration of the parameter space and improves the robustness of the fit.

For a $J^P=1/2^-$ intermediate $\Sigma$ resonance,
\begin{align}
    \mathcal{L}_{KN\Sigma(1/2^-)} &= -ig_{KN\Sigma(1/2^-)} \bar{K}\, \bar{\Sigma}\left(1/2^-\right) \cdot \tau N + \mathrm{H.c.},\\
    \mathcal{L}_{\pi\Sigma\Sigma(1/2^-)} &= g_{\pi\Sigma\Sigma(1/2^-)} \bar{\Sigma}\left(1/2^-\right) \times \Sigma \cdot\pi + \mathrm{H.c.},\\
    \mathcal{L}_{\pi\Lambda\Sigma(1/2^-)} &=-ig_{\pi\Lambda\Sigma(1/2^-)}\bar{\Sigma}\left(1/2^-\right) \Lambda\pi+\mathrm{H.c.}.
\end{align}
The products $g_{KN\Sigma(1/2^-)}g_{\pi\Sigma\Sigma(1/2^-)}$
and $g_{KN\Sigma(1/2^-)}g_{\pi\Lambda\Sigma(1/2^-)}$ are constrained by
the magnitudes estimated from the corresponding decay widths.

For a $J^P=3/2^+$ intermediate $\Sigma^*$ resonance,
\begin{align}
    \mathcal{L}_{KN\Sigma^*} &= \frac{f_{KN\Sigma^*}}{m_K} \partial_\mu \bar{K}\, \bar{\Sigma}^{*\mu} \cdot \tau N + \mathrm{H.c.},\\
    \mathcal{L}_{\pi\Sigma\Sigma^*} &= i\frac{f_{\pi\Sigma\Sigma^*}}{m_\pi} \partial_\mu \pi\cdot \bar{\Sigma}^{*\mu} \times\Sigma + \mathrm{H.c.},\\
    \mathcal{L}_{\pi\Lambda\Sigma^*} &=\frac{f_{\pi\Lambda\Sigma^*}}{m_\pi}\partial_\mu \pi\cdot\bar{\Sigma}^{*\mu}\Lambda+\mathrm{H.c.}.
\end{align}
For $\Sigma(1385)$, $f_{\pi\Sigma\Sigma^\ast}=0.68$ and $f_{\pi\Lambda\Sigma^\ast}=1.27$ follow from the partial widths $\Gamma_{\Sigma^\ast\to\Sigma\pi}\simeq4.2$~MeV and $\Gamma_{\Sigma^\ast\to\Lambda\pi}\simeq31$~MeV \cite{ParticleDataGroup:2024cfk}. They are close to the SU(3) estimates 0.89 and 1.54~\cite{Doring:2010ap}. We take $f_{KN\Sigma^\ast}=-3.22$ from SU(3) flavor symmetry ~\cite{Oh:2007jd} and allow both coupling products to vary between one half and twice their central values.

For a $J^P=3/2^-$ intermediate state,

\begin{align}
    \mathcal{L}_{KN\Sigma(3/2^-)} &=\frac{f_{KN\Sigma(3/2^-)}}{m_K} \partial_\mu \bar{K}\, \bar{\Sigma}^\mu\left(3/2^-\right) \cdot\tau\gamma_5N+\mathrm{H.c.},\\
    \mathcal{L}_{\pi\Sigma\Sigma(3/2^-)} &= i\frac{f_{\pi \Sigma\Sigma(3/2^-)}}{m_\pi}\partial_\mu\pi\cdot \bar{\Sigma}^\mu\left(3/2^-\right) \times\gamma_5 \Sigma +\mathrm{H.c.},\\
    \mathcal{L}_{\pi\Lambda\Sigma(3/2^-)} &=\frac{f_{\pi\Lambda\Sigma(3/2^-)}}{m_\pi}\partial_\mu\pi\bar{\Sigma}^\mu\left(3/2^-\right) \gamma_5\Lambda+\mathrm{H.c.}.
\end{align}
The central values inferred from the $\Sigma(1670)\,3/2^-$ widths are
\begin{equation}
    \frac{f_{KN\Sigma(3/2^-)}}{m_K}=5.2,\;
    \frac{f_{\pi \Sigma\Sigma(3/2^-)}}{m_\pi}=15.7,\;
    \frac{f_{\pi \Lambda\Sigma(3/2^-)}}{m_\pi}=7.4
\end{equation}
Because the relevant partial widths are uncertain, each product is varied by
an overall factor between one half and two.

For a $J^P=5/2^-$ intermediate state,
\begin{align}
    \mathcal{L}_{KN\Sigma(5/2^{-})} &=g_{KN\Sigma(5/2^{-})}\partial_{\mu}\partial_{\nu}\bar{K}\, \bar{\Sigma}^{\mu\nu}\left(5/2^-\right)\cdot\tau N +\mathrm{H.c.},\\
    \mathcal{L}_{\pi\Sigma\Sigma(5/2^{-})} &=ig_{\pi\Sigma\Sigma(5/2^{-})}\partial_{\mu}\partial_{\nu}\pi \cdot \bar{\Sigma}^{\mu\nu}\left(5/2^-\right)\times \Sigma +\mathrm{H.c.},\\
    \mathcal{L}_{\pi\Lambda\Sigma(5/2^-)} &=g_{\pi\Lambda\Sigma(5/2^-)}\partial_\mu\partial_\nu\pi\cdot\bar{\Sigma}^{\mu\nu}\left(5/2^-\right) \Lambda+\mathrm{H.c.}
\end{align}
The values inferred from the $\Sigma(1775)\,5/2^-$ widths are
\begin{equation}
    g_{KN\Sigma(5/2^-)}=7.7,\;
    g_{\pi \Sigma\Sigma(5/2^-)}=2.4,\;
    g_{\pi \Lambda\Sigma(5/2^-)}=4.3
\end{equation}
and the two products are again varied by a factor between one half and two.

\subsection{u-channel nucleon exchange}

The $u$-channel terms use Eq.~\eqref{S-eq:KNSigma} together with
\begin{align}
    \mathcal{L}_{\pi NN} &=\frac{g_{\pi NN}}{2M_N}\bar{N}\gamma^\mu\gamma_5\partial_\mu\pi\cdot\tau N,\\
    \mathcal{L}_{KN\Lambda} &=\frac{g_{KN\Lambda}}{M_N+M_\Lambda}\bar{N}\gamma^\mu\gamma_5\Lambda\partial_\mu K+\mathrm{H.c.},
\end{align}
where
\begin{equation}
    g_{\pi NN}=13.26,\qquad g_{KN\Lambda}=-13.24,
\end{equation}
as determined from SU(3) flavor symmetry~\cite{Oh:2006hm,Oh:2004wp,Gao:2010ve,Arndt:2006bf}.

\subsection{t-channel $K^*$ exchange and form factors}

For $K^*$ exchange, we employ
\begin{align}
    \mathcal{L}_{K^*K\pi}&=ig_{K^*K\pi}\bar{K}_\mu^*\left(\pi\cdot\tau\partial^\mu K-\partial^\mu\pi\cdot\tau K\right),\\
    \mathcal{L}_{K^*N\Sigma}&=-g_{K^*N\Sigma}\bar{\Sigma}\cdot\tau \left(\gamma_\mu \bar{K}^{*\mu}-\frac{\kappa_{K^*N\Sigma}}{2M_N}\sigma_{\mu\nu}\partial^\nu \bar{K}^{*\mu}\right)N + \mathrm{H.c.},\\
    \mathcal{L}_{K^*N\Lambda}&=-g_{K^*N\Lambda}\bar{\Lambda}\left(\gamma_\mu K^{*\mu}-\frac{\kappa_{K^*N\Lambda}}{2M_N}\sigma_{\mu\nu}\partial^\nu K^{*\mu}\right)N+\mathrm{H.c.}
\end{align}
The value $g_{K^* K\pi}=-3.23$ is determined from $K^*\to K\pi$, with its sign fixed by SU(3) relations~\cite{Mueller-Groeling:1990uxr}. Two SU(3)-based Nijmegen soft-core estimates, NSC97a and NSC97f, are \cite{Stoks:1999bz,Oh:2006hm,Oh:2004wp,Shi:2014vha}
\begin{align}
    g_{K^\ast N\Sigma}&=-2.46,& \kappa_{K^\ast N\Sigma}&=-0.47 &&(\mathrm{NSC97a}),\\
    g_{K^\ast N\Sigma}&=-3.52,& \kappa_{K^\ast N\Sigma}&=-1.14 &&(\mathrm{NSC97f}),\\
    g_{K^\ast N\Lambda}&=-4.26,& \kappa_{K^\ast N\Lambda}&=2.66 &&(\mathrm{NSC97a}),\\
    g_{K^\ast N\Lambda}&=-6.11,& \kappa_{K^\ast N\Lambda}&=2.43 &&(\mathrm{NSC97f}).
\end{align}
The fit ranges are
\begin{align}
    g_{K^\ast N\Sigma}&\in[-7.0,-1.2],& \kappa_{K^\ast N\Sigma}&\in[-2.3,-0.2],\\
    g_{K^\ast N\Lambda}&\in[-12.2,-2.1],& \kappa_{K^\ast N\Lambda}&\in[1.2,5.3],
\end{align}
to incorporate theoretical uncertainties.

At each hadronic vertex we introduce the dipole form factor
\begin{equation}
    F_B(q_{\rm ex}^2,M_{\rm ex}) =\frac{\Lambda^4}{\Lambda^4+(q_{\rm ex}^2-M_{\rm ex}^2)^2},
\end{equation}
where $q_{\rm ex}$ and $M_{\rm ex}$ are the momentum and mass of the exchanged
particle.  Each cutoff $\Lambda$ is restricted to $0.5$--$2.0$~GeV.

\section*{S2: Propagators, amplitudes, and observables}

The $t$-channel vector-meson and $u$-channel nucleon propagators are
\begin{align}
    G_{K^\ast}^{\mu\nu}(p) &=\frac{-g^{\mu\nu}+p^\mu p^\nu/m_{K^\ast}^2}{p^2-m_{K^\ast}^2},\\
    G_B(q)&=\frac{\slashed q+M_N}{q^2-M_N^2}.
\end{align}
For spin-$1/2$, $3/2$, and $5/2$ $s$-channel resonances we use
\begin{align}
    G_R^{1/2}(q) &=\frac{\slashed q+M}{q^2-M^2+iM\Gamma},\\
    G_R^{3/2,\mu\nu}(q) &=\frac{\slashed q+M}{q^2-M^2+iM\Gamma}\left[-g^{\mu\nu}+\frac{\gamma^\mu\gamma^\nu}{3}\right.\nonumber\\
    &\hspace{1.2cm}\left. +\frac{\gamma^\mu q^\nu-\gamma^\nu q^\mu}{3M} +\frac{2q^\mu q^\nu}{3M^2}\right],\\
    G_R^{5/2,\alpha\beta\mu\nu}(q)&=\frac{\slashed q+M}{q^2-M^2+iM\Gamma}S_{\alpha\beta\mu\nu}(q,M),
\end{align}
where
\begin{align}
    S_{\alpha\beta\mu\nu}(q,M)=&\frac{1}{2}(\bar{g}_{\alpha\mu}\bar{g}_{\beta\nu}+\bar{g}_{\alpha\nu}\bar{g}_{\beta\mu})
    -\frac{1}{5}\bar{g}_{\alpha\beta}\bar{g}_{\mu\nu}-\frac{1}{10}(\bar{\gamma}_\alpha\bar{\gamma}_\mu\bar{g}_{\beta\nu} \nonumber\\
    &+\bar{\gamma}_\alpha\bar{\gamma}_\nu\bar{g}_{\beta\mu}+\bar{\gamma}_\beta\bar{\gamma}_\mu\bar{g}_{\alpha\nu}
    +\bar{\gamma}_\beta\bar{\gamma}_\nu\bar{g}_{\alpha\mu}),\\
    \bar g_{\mu\nu}&=g_{\mu\nu}-\frac{q_\mu q_\nu}{M^2},\\
    \bar\gamma_\mu&=\gamma_\mu-\frac{q_\mu\slashed q}{M^2}.
\end{align}

For either $\bar{K}^0(k_1) +p(p_1)\to\pi^+(k_2) +\Sigma^0/\Lambda(p_2)$, the scattering amplitude is
\begin{align}
    \mathcal{M}_{\bar{K}^0p\to\pi^+\Sigma^0/\Lambda}^{r_2,r_1}&=\bar{u}_{r_2}(p_2)\mathcal{A}u_{r_1}(p_1) \nonumber\\ 
    &=\bar{u}_{r_2}(p_2)\left(\sum_i\mathcal{A}_i\right)u_{r_1}(p_1),
\end{align}
where $r_1$ and $r_2$ denote the polarizations of the initial nucleon and the final $\Sigma^0$ or $\Lambda$, respectively, and $\mathcal{A}_i$ denotes the non-spinor part of the $i$th diagram. 

The differential cross section is
\begin{align}
    \frac{\textmd{d}\sigma_{K_Lp\to\pi^+\Sigma^0/\Lambda}}{\textmd{d}\Omega}&=\frac{\textmd{d}\sigma_{K_Lp\to\pi^
    +\Sigma^0/\Lambda}}{2\pi\, \textmd{d} \cos(\theta)} \nonumber\\ 
    &=\frac{1}{2}\frac{1}{64\pi^2s}\frac{|\vec{k}_2|}{|\vec{k}_1|}\overline{|\mathcal{M}_{\bar{K}^0p\to
    \pi^+\Sigma^0/\Lambda}|}^2,
\end{align}
where the first factor $1/2$ originates from the $K_L\to\bar K^0$ projection. 
The spin-averaged squared amplitude is
\begin{align}
    \overline{|\mathcal{M}_{\bar{K}^0p\to\pi^+\Sigma^0/\Lambda}|}^2 &=\frac{1}{2}\sum_{r_1,r_2}\mathcal{M}_{\bar{K}^0p\to\pi^+\Sigma^0/\Lambda}^{r_2,r_1} \mathcal{M}_{\bar{K}^0p\to\pi^+\Sigma^0/\Lambda}^{\dagger r_2,r_1} \nonumber\\
    &=\frac{1}{2}\mathrm{Tr}\left[(\slashed{p}_2+m_{\Sigma/\Lambda})\mathcal{A} (\slashed{p}_1+m_N) \gamma^0\mathcal{A}^\dagger\gamma^0 \right].
\end{align}

For an unpolarized beam and target, the recoil polarization of the $\Sigma^0$ and $\Lambda$ normal to the reaction plane are~\cite{Amaryan:2016ufk,Shi:2014vha,Penner:2002ma}
\begin{equation}
    P_{\Sigma/\Lambda}=-\frac{2\mathrm{Im}{\left( \mathcal{M}_{\bar{K}^0p\to\pi^+\Sigma^0/\Lambda}^{1/2,1/2} \mathcal{M}_{\bar{K}^0p\to\pi^+\Sigma^0/\Lambda}^{*-1/2,1/2}\right)} } {\overline{|\mathcal{M}_{\bar{K}^0p\to\pi^+\Sigma^0/\Lambda}|}^2},
\end{equation}
which measures the asymmetry of the recoil spin distribution along $\hat{\mathbf{v}}=\hat{\mathbf{k}}_1\times\hat{\mathbf{k}}_2$, i.e. normal to the reaction plane.

\section*{S3: Complete fit parameters and alternative fits}

Using the differential cross sections and recoil polarizations defined in Sec. S1 and S2, with all couplings constrained within physically motivated ranges, we perform the first combined analysis of the two isospin-selective reactions $K_L p\to \pi^+\Sigma^0$ and $K_L p\to \pi^+\Lambda$, incorporating theoretically complete tree-level dynamics. 

The nonresonant $u$- and $t$-channel terms and the four-star
$\Sigma(1189)$, $\Sigma(1385)$, $\Sigma(1670)$, and $\Sigma(1775)$ states are
retained in every scenario.  Their removal leads to a visibly poorer
description.  With only these contributions, the 18-parameter baseline fit
gives $\chi^2/\mathrm{d.o.f.}=5.161$.  The optimal solution further contains
$\Sigma(1580)$, $\Sigma(1620)$, $\Sigma(1660)$, and $\Sigma(1750)$.

\renewcommand{\arraystretch}{1.5}  
\begin{table*}[!tbph]
    \centering
    \caption{Complete fitted parameter set. Optimal fit denotes the solution used in the main text. Fit I removes $\Sigma(1750)\,1/2^-$; Fit II fixes the $\Sigma(1620)\,1/2^-$ mass to 1.4~GeV; Fit III adds only $\Sigma(1620)\,1/2^-$ to the baseline. Bracketed ranges indicate estimates based on SU(3) symmetry relations or PDG-reported values~\cite{ParticleDataGroup:2024cfk}. A dash (---) denotes parameters absent from a fit scenario.}
    \label{S-tab:full}
    \setlength{\tabcolsep}{8pt} 
    \begin{tabular}{l|llllll}
    \toprule[1.50pt]
    \toprule[0.50pt]
    Resonances & Parameters & Optimal Fit & Fit I & Fit II & Fit III & Estimates \\ \hline
    $K^*$    & $g_{K^*N\Sigma}$      & -1.2$\pm$0.1   & -1.4  & -2.4& -1.8  & [-7.0, -1.2]\\
             & $\kappa_{K^*N\Sigma}$ & -2.3$\pm$0.2   & -2.3  & -0.2& -2.3  & [-2.3, -0.2]\\
             & $g_{K^*N\Lambda}$     & -9.8$\pm$0.1   & -4.1  & -7.9& -5.2 & [-12.2, -2.1]\\
             & $\kappa_{K^*N\Lambda}$&  1.2$\pm$0.1   &  1.2  & 1.76& 1.2 & [1.2, 5.3]\\
             & $\Lambda$   & 1.14$\pm$0.01 & 2.0 & 1.1  & 1.4  & [0.5, 2.0] \\ \hline
             
    $N$      & $\Lambda$   & 0.97$\pm$0.01 & 0.94 & 1.1 & 1.0    & [0.5, 2.0] \\ \hline
    
    $\Sigma(1189) 1/2^+$   & $g_{KN\Sigma}f_{\pi\Sigma\Sigma}$ & -1.35$\pm$0.04 & -1.50 & -1.35 & -1.50 & [-5.4, -1.3] \\
            & $g_{KN\Sigma}f_{\pi\Lambda\Sigma}$  & 69.6$\pm$2.8 & 66.0 & 47.9& 17.4 & [17.4, 69.6]\\
            & $\Lambda$    & 1.01$\pm$0.02   & 0.50 & 0.51 & 0.5  & [0.5, 2.0] \\ \hline
                           
    $\Sigma(1385) 3/2^+$  & $f_{KN\Sigma^*} f_{\pi\Sigma\Sigma^*}$  & -1.12$\pm$0.10 & -1.12 & -1.12 & -1.12 & [-5.7, -1.1] \\
            & $f_{KN\Sigma^*} f_{\pi\Lambda\Sigma^*}$ & -8.63$\pm$0.54 & -8.63 & -5.35& -8.63& [-8.6, -2.1]\\
            & $\Lambda$     & 0.61$\pm$0.01 & 0.66 & 0.79 & 0.66   & [0.5, 2.0] \\ \hline
                          
    $\Sigma(1670) 3/2^-$  & $\sqrt{\Gamma_{\bar{K}N}\Gamma_{\pi\Sigma} }/\Gamma_{tot}$  & +0.20$\pm$0.01 & +0.18& +0.12 & +0.18 & [0.09, 0.38] \\
            & $\sqrt{\Gamma_{\bar{K}N}\Gamma_{\pi\Lambda} }/\Gamma_{tot}$ & +0.14$\pm$0.01 & +0.05 & +0.16 & +0.06 & [0.04, 0.18] \\
            & $\Lambda$     & 0.97$\pm$0.12 & 2.0 & 0.95 & 2.0 & [0.5, 2.0] \\ \hline
                          
    $\Sigma(1775) 5/2^-$  & $\sqrt{\Gamma_{\bar{K}N}\Gamma_{\pi\Sigma} }/\Gamma_{tot}$  & +0.24$\pm$0.02 & +0.24& +0.24 & +0.24 & [0.06, 0.24] \\
            & $\sqrt{\Gamma_{\bar{K}N}\Gamma_{\pi\Lambda} }/\Gamma_{tot}$ & -0.15$\pm$0.02 & -0.43 & -0.13& -0.40& [-0.52, -0.13]\\
            & $\Lambda$     & 2.00$\pm$0.23   & 2.0 & 2.0 & 1.37   & [0.5, 2.0] \\ \hline
                          
    $\Sigma(1580) 3/2^-$  & $\sqrt{\Gamma_{\bar{K}N}\Gamma_{\pi\Sigma} }/\Gamma_{tot}$  & +0.010$\pm$0.004 & +0.004& +0.012 & --- & [-0.35, 0.35] \\
            & $\sqrt{\Gamma_{\bar{K}N}\Gamma_{\pi\Lambda} }/\Gamma_{tot}$ & -0.012$\pm$0.002 & -0.002& -0.005& --- & [-0.39, 0.39]\\
            & $\Lambda$     & 0.50$\pm$0.11 & 0.50 & 0.50 & --- & [0.5, 2.0] \\ \hline
                          
    $\Sigma(1660) 1/2^+$  & $M$ [GeV] & 1.637$\pm$0.003 &  1.613  & 1.75 & --- & [1.40, 1.75] \\
                          & $\Gamma$ [GeV] & 0.130$\pm$0.006 & 0.105& 0.01 & --- & [0.01, 0.40] \\
                          & $\sqrt{\Gamma_{\bar{K}N}\Gamma_{\pi\Sigma} }/\Gamma_{tot}$  & -0.052$\pm$0.005 & +0.001 & +0.112 & --- & [-0.48, 0.48] \\
            & $\sqrt{\Gamma_{\bar{K}N}\Gamma_{\pi\Lambda} }/\Gamma_{tot}$ & -0.077$\pm$0.002 & -0.055 & +0.086& --- & [-0.41, 0.41]\\
            & $\Lambda$     & 2.0$\pm$1.4   & 2.0 & 1.8 & --- & [0.5, 2.0] \\ \hline
                          
    $\Sigma(1620) 1/2^-$  & $M$ [GeV] & 1.557$\pm$0.002 & 1.556 & 1.4(Fixed) & 1.554 & [1.35, 1.65] \\
                          & $\Gamma$ [GeV] & 0.117$\pm$0.001 & 0.086  & 0.10 & 0.086 & [0.01, 0.40] \\
                          & $\sqrt{\Gamma_{\bar{K}N}\Gamma_{\pi\Sigma} }/\Gamma_{tot}$  & -0.741$\pm$0.006 & -0.50 & -1.68 & -0.51 & [-3.2, 3.2] \\
            & $\sqrt{\Gamma_{\bar{K}N}\Gamma_{\pi\Lambda} }/\Gamma_{tot}$ & -0.142$\pm$0.006 & -0.96 & +0.49& +0.003 & [-2.4, 2.4]\\
            & $\Lambda$     & 0.62$\pm$0.01 & 0.83 & 2.0 & 0.9 & [0.5, 2.0] \\ \hline
                          
    $\Sigma(1750) 1/2^-$  & $\sqrt{\Gamma_{\bar{K}N}\Gamma_{\pi\Sigma} }/\Gamma_{tot}$  & +0.45$\pm$0.01 & --- & --- & --- & [-1.2, 1.2] \\
            & $\sqrt{\Gamma_{\bar{K}N}\Gamma_{\pi\Lambda} }/\Gamma_{tot}$ & +0.19$\pm$0.01 & --- & --- & --- & [-1.2, 1.2]\\
            & $\Lambda$     & 2.0$\pm$1.2 & --- & --- & ---   & [0.5, 2.0] \\ \hline
                    & \text{D.o.F}           & 533    & 536    & 537   & 544     & \\
                    & $\chi^2/\text{D.o.F}$  & 1.604  & 2.210  & 1.880 & 2.546   & \\
    \bottomrule[0.50pt]
    \bottomrule[1.50pt]
    \end{tabular}
\end{table*}
\setlength{\extrarowheight}{0pt}

The optimal solution summarized in Table~\ref{S-tab:full} gives a good simultaneous description of all available angular distributions and recoil-polarization data, illustrating the stability and discriminating power of the present multichannel analysis.

Fit I shows that $\Sigma(1750)\,1/2^-$ becomes important when both channels are described simultaneously: removing it increases $\chi^2/\mathrm{d.o.f.}$ from 1.604 to 2.210, although it was not strongly required in the earlier single-channel analysis~\cite{Guo:2025mha}. 
In Fit II, fixing the $\Sigma(1620)$ mass to 1.4~GeV worsens the fit to 1.880 and drives the $\Sigma(1660)$ mass and width to the edges of their allowed ranges.  
Fit III shows that $\Sigma(1620)$ alone accounts for a substantial part of the missing dynamics, reducing the baseline result from 5.161 to 2.546; a satisfactory global description nevertheless requires the additional $\Sigma(1580)$, $\Sigma(1660)$, and $\Sigma(1750)$ contributions. 


\section*{S4: Complete angular distributions and polarizations}

\begin{figure*}[!htbp]
    \centering
    \includegraphics[width=0.9\textwidth]{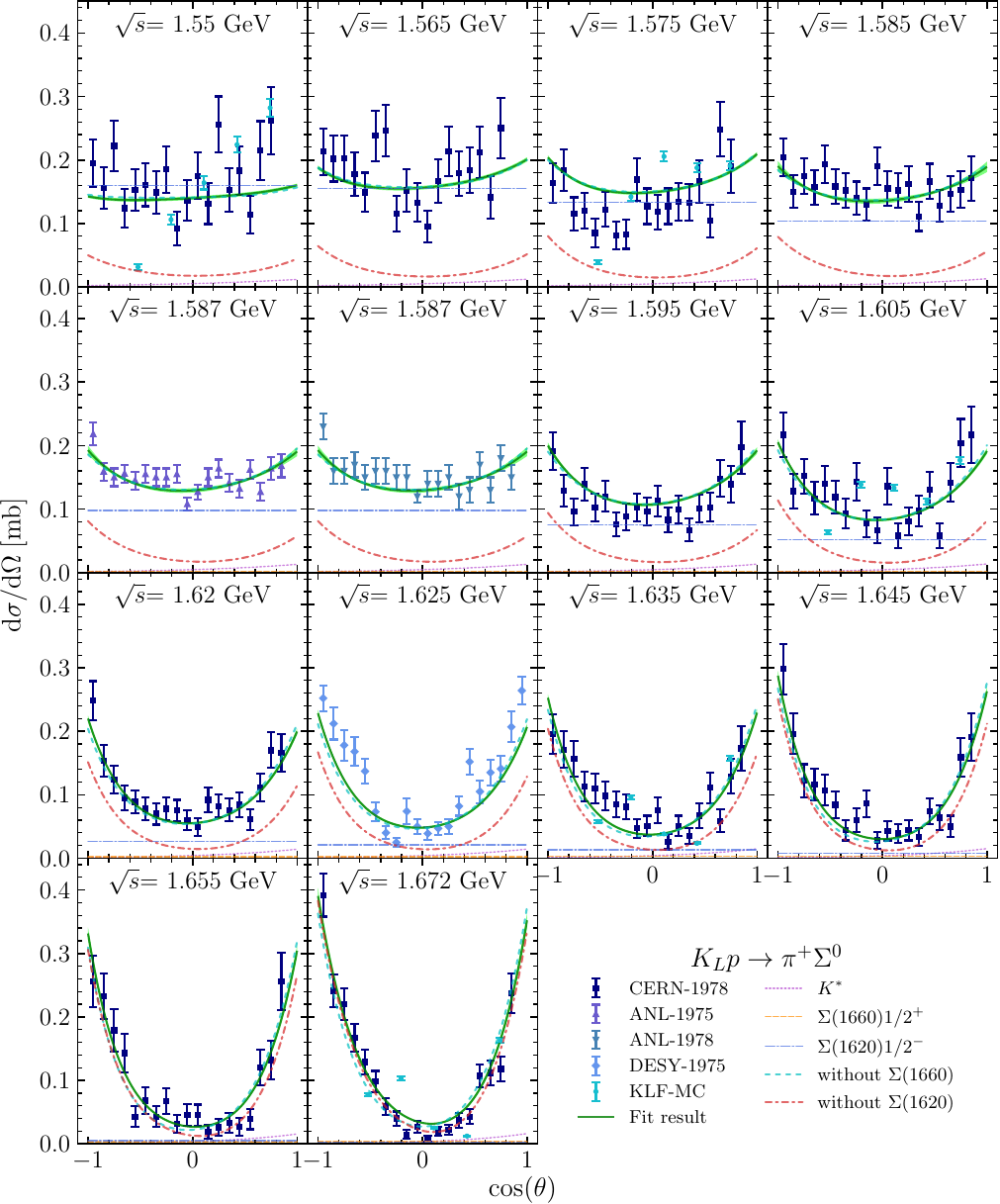}
    \captionof{figure}{Differential cross sections for $K_Lp\to\pi^+\Sigma^0$ over the full fitted energy range.  The optimal result and its $1\sigma$ fit band are shown together with the isolated $K^*$, $\Sigma(1660)$, and $\Sigma(1620)$ contributions and the results obtained after removing either resonance.
    The measurements are from CERN-1978~\cite{Bologna-Edinburgh-Glasgow-Pisa-Rutherford:1977cyz}, ANL-1975~\cite{Cho:1975dv}, ANL-1978~\cite{Engler:1978rr} and DESY-1975~\cite{Burkhardt:1975sd}. KLF Monte Carlo points were not fitted.}
    \label{S-fig:dcsSigma}
\end{figure*}

\begin{figure*}
    \centering
    \includegraphics[width=0.9\textwidth]{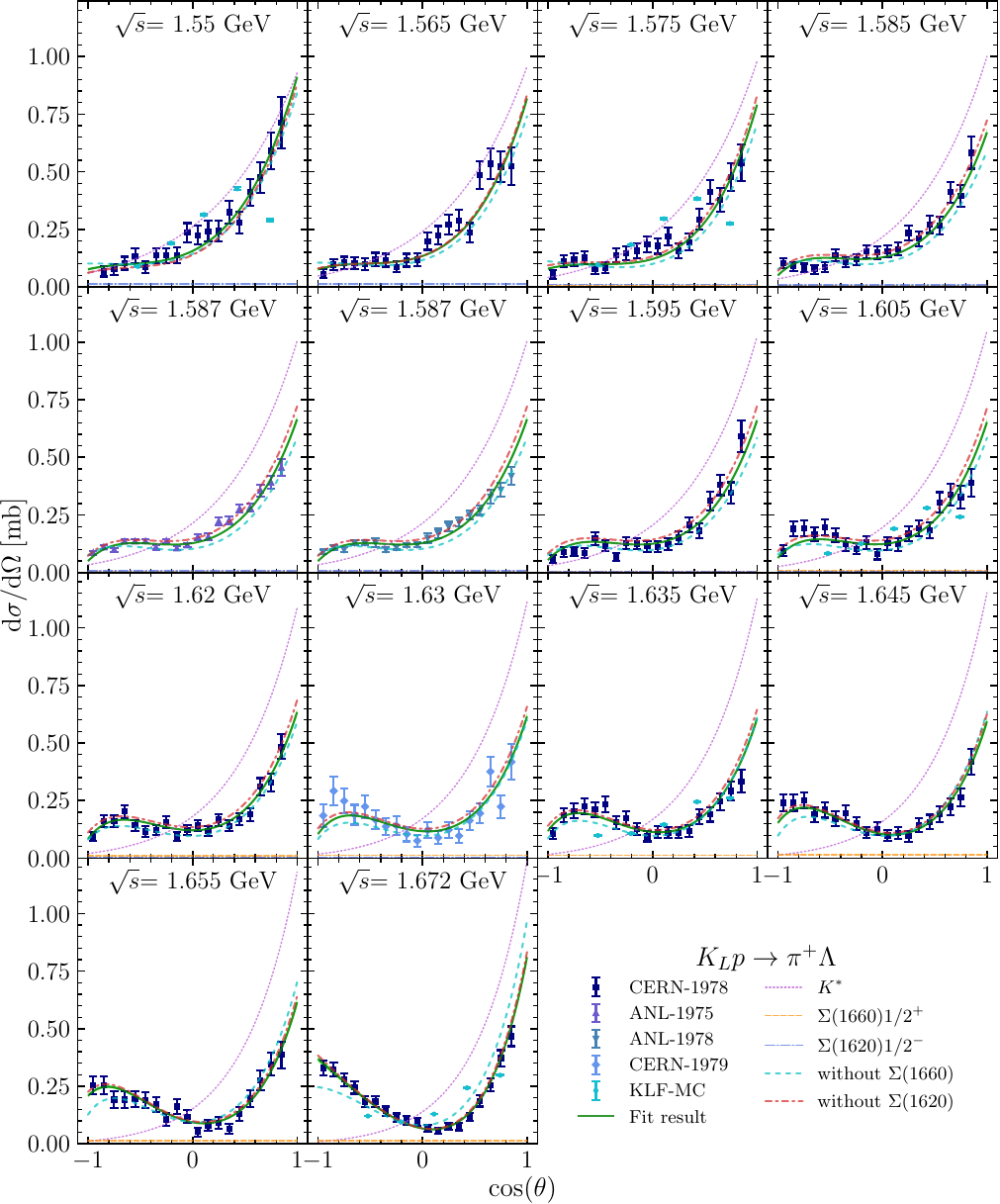}
    \caption{Differential cross sections for $K_Lp\to\pi^+\Lambda$ over the full fitted energy range, with the same curve conventions as in Fig.~\ref{S-fig:dcsSigma}. The measurements are from CERN-1978~\cite{Bologna-Edinburgh-Glasgow-Pisa-Rutherford:1977cyz}, ANL-1975~\cite{Cho:1975dv}, ANL-1978~\cite{Engler:1978rr}, and CERN-1979~\cite{Corden:1979tz}. KLF Monte Carlo points were not fitted.}
    \label{S-fig:dcsLambda}
\end{figure*}

\begin{figure*}
\centering
    \includegraphics[width=0.9\textwidth]{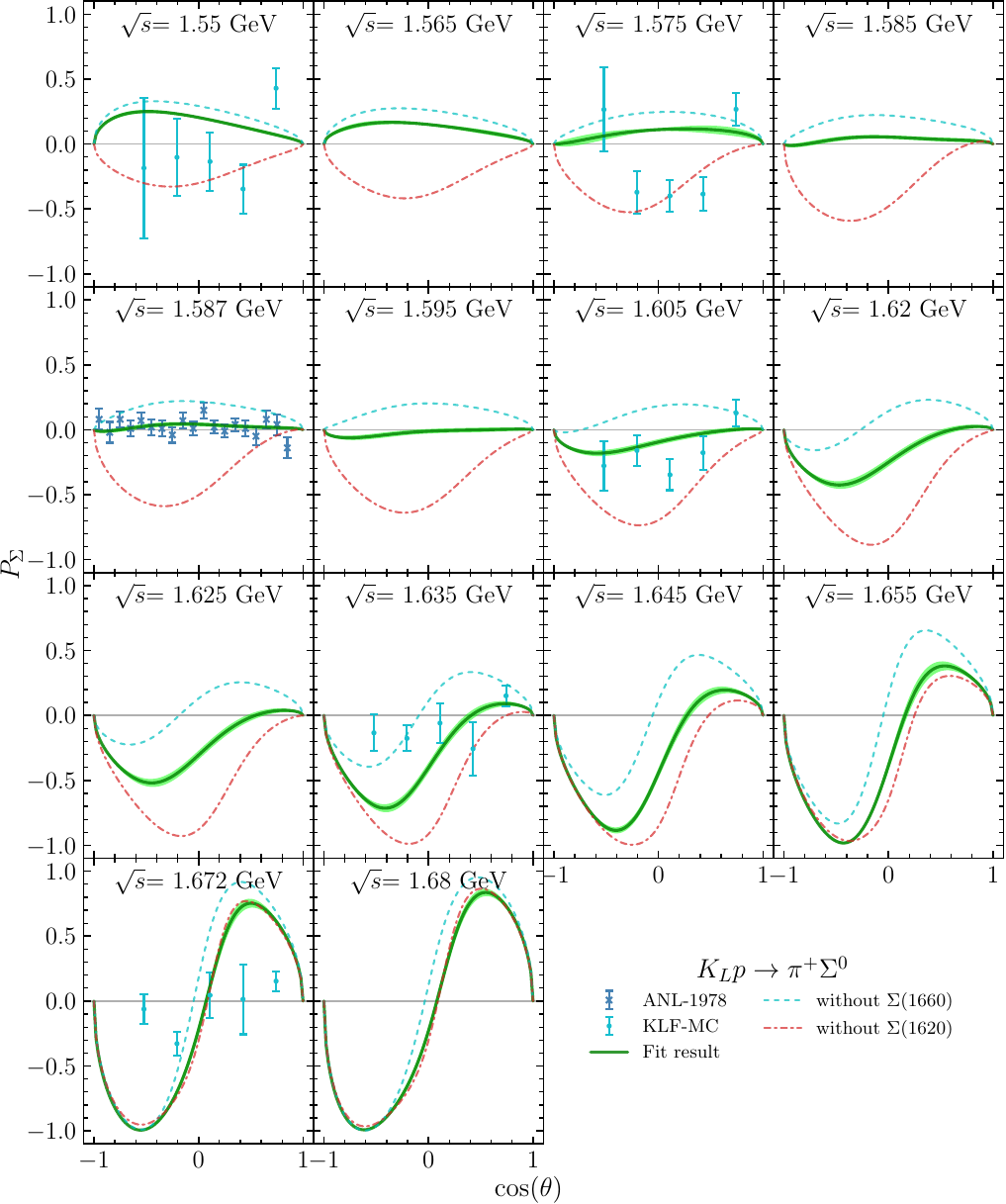}
    \caption{Recoil polarization of the $\Sigma^0$ in $K_Lp\to\pi^+\Sigma^0$.  Results without $\Sigma(1660)$ or $\Sigma(1620)$ are shown with the optimal fit and its $1\sigma$ band.  The data at $\sqrt{s}=1.587$~GeV are from ANL-1978~\cite{Engler:1978rr}. KLF Monte Carlo points were not fitted.}
    \label{S-fig:polSigma}
\end{figure*}

\begin{figure*}
    \centering
    \includegraphics[width=0.9\textwidth]{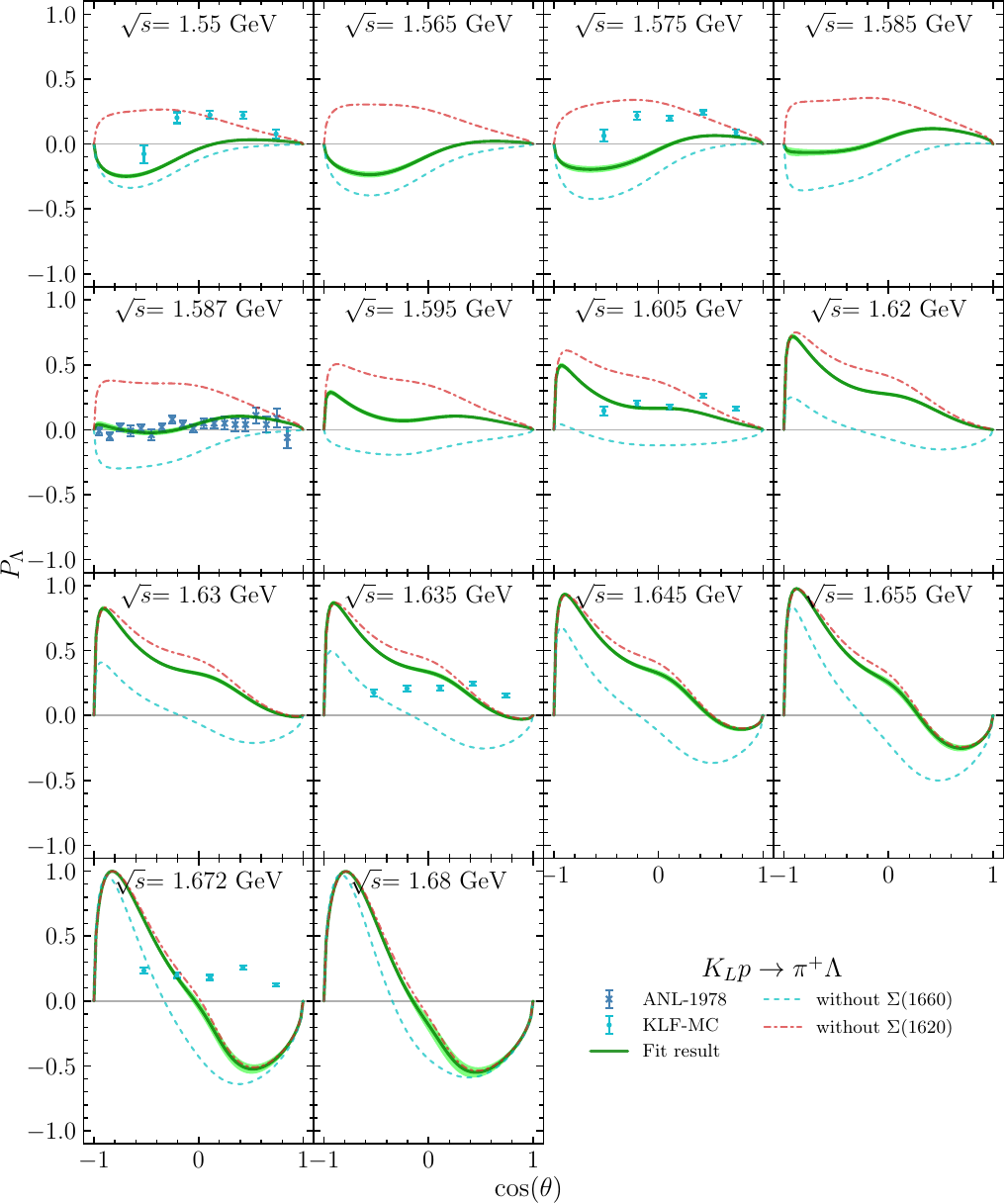}
    \captionof{figure}{Recoil polarization of the $\Lambda$ in $K_Lp\to\pi^+\Lambda$, with the same conventions as Fig.~\ref{S-fig:polSigma}.  The data at $\sqrt{s}=1.587$~GeV are from ANL-1978~\cite{Engler:1978rr}.  KLF Monte Carlo points were not fitted.}
    \label{S-fig:polLambda}
\end{figure*}

Figures~\ref{S-fig:dcsSigma}--\ref{S-fig:polLambda} show the full energy dependence underlying the selected comparisons in the main text.  The KLF Monte Carlo
points in these figures are prospective projections and were not included in
the fit.

The Monte Carlo package developed for the KLF simulates neutral-kaon interactions with liquid $\mathrm{H}_2$ and $\mathrm{D}_2$ targets. Final states are generated using measured kaon-scattering cross sections and propagated through a \textsf{Geant4} simulation of the GlueX detector with the KLF target configuration. The detector response is digitized, smeared, and reconstructed using the standard Hall~D/GlueX software, extended to reconstruct the neutral-kaon beam. Individual reaction channels are selected using time-of-flight and drift-chamber ionization measurements, together with kinematic fits imposing the appropriate primary- or secondary-vertex constraints. 

Figure~\ref{S-fig:dcsSigma} shows the fitted differential cross sections for $K_L p\to \pi^+\Sigma^0$. The displayed single-diagram contributions show that the $t$-channel $K^*$ exchange is responsible for the forward-angle rise, whereas the $s$-channel resonance terms mainly reshape the distributions through interference. In this channel, the direct contribution of $\Sigma(1620) 1/2^-$ is comparatively prominent, especially at the lower and intermediate energies, and removing it leads to a clear deterioration of the angular distributions. The direct contribution of $\Sigma(1660) 1/2^+$ is smaller, but the comparison with the ``without $\Sigma(1660)$'' curves shows that it still plays a non-negligible role through interference with the background and with the other $s$-channel states. Therefore, the $\pi^+\Sigma^0$ channel remains the most sensitive probe of the $\Sigma(1620)$ region. 

The situation is noticeably different in Fig.~\ref{S-fig:dcsLambda} for $K_L p\to \pi^+\Lambda$. The angular distributions in this channel are much more strongly forward peaked, indicating that the $t$-channel $K^*$ exchange is more important in $\pi^+\Lambda$ than in $\pi^+\Sigma^0$. 
In other words, although the same basic mechanisms are included in both reactions, their relative importance is channel dependent. 
This is one of the central physical messages of the present multichannel analysis: a contribution that is subleading in one channel may become dominant in the other, and only a combined fit can consistently determine their relative weights. 
The direct effect of $\Sigma(1620) 1/2^-$ on the differential cross section is less pronounced here than in the $\pi^+\Sigma^0$ channel, while the role of $\Sigma(1660) 1/2^+$ is relatively more visible. 
This behavior is in line with our previous analyses of the $\pi\Lambda$ channel~\cite{Gao:2010ve,Gao:2012zh}, where the $\Sigma(1620)$ contribution was found to be unnecessary. The present result therefore does not contradict that conclusion; rather, it shows that once the $\pi^+\Lambda$ data are analyzed together with the more $\Sigma(1620)$-sensitive $\pi^+\Sigma^0$ channel under common phase and parameter constraints, the $\Sigma(1620)$ term becomes necessary in the global fit, reflecting the complementarity of the two channels.

Figures~\ref{S-fig:polSigma} and \ref{S-fig:polLambda} display the recoil polarizations for the $\Sigma^0$ and $\Lambda$, respectively. These observables are especially important because they arise from the interference among different amplitudes and are therefore much more sensitive to relative phases than the differential cross sections alone.  A clear evolution is observed from weak angular dependence near threshold to strong forward-backward structures at higher energies. 
It is worth emphasizing again that the uncertainty bands in Fig.~\ref{S-fig:polSigma} are very narrow. In the previous single-channel study~\cite{Guo:2025mha, Gao:2010ve, Gao:2012zh}, the recoil-polarization bands were visibly broader because the fit was driven mainly by the differential cross sections. Here, however, even the polarization predictions are very stable, directly reflecting the additional constraints brought in by the simultaneous fit to the $\pi^+\Lambda$ channel. 

The $\Lambda$ recoil polarization shown in Fig.~\ref{S-fig:polLambda} exhibits a markedly different angular pattern from that of the $\Sigma^0$ polarization. This channel-dependent behavior again confirms that the two final states probe different combinations of partial waves and different background-resonance interference mechanisms. Therefore, the two polarization observables are not redundant. In particular, the simultaneous description of $P_\Sigma$ and $P_\Lambda$ strongly suppresses parameter sets that would otherwise remain allowed if only one final state were considered.

\twocolumngrid

\end{document}